\documentclass[aps,amsmath,twocolumn,amssymb,floatfixng,showpacs,
superscriptaddress,footinbib]{revtex4-1}

\usepackage{ulem}
\usepackage{graphicx,epstopdf}  
\usepackage{amsmath}
\usepackage{graphicx}
\usepackage{amsfonts}
\usepackage{color}
\usepackage{dcolumn}   
\usepackage{bm}        
\usepackage{amssymb}   
\usepackage{hyperref}
\begin{document}

\title{Dynamics of fluctuation correlation in a periodically driven classical system}
\author{Aritra Kundu}\email{akundu@sissa.it}\affiliation{SISSA and INFN, via Bonomea 265, 34136 Trieste, Italy }
\author{Atanu Rajak}\email{raj.atanu009@gmail.com}\affiliation{Presidency University, 86/1, College Street, Kolkata 700073, India}
\author{Tanay Nag}\email{tnag@physik.rwth-aachen.de}\affiliation{Institute f\"ur Theorie der Statistischen Physik, RWTH Aachen University, 52056 Aachen, Germany}
 
\date{\today}

\def\nn{\nonumber}
\def\redw#1{{\color{red} #1}}
\def\greenw#1{{\color{dgreen} #1}}
\def\bluew#1{{\color{blue} #1}}
\def\la{\langle}
\def\ra{\rangle}
\def\ql{ {Q_{ l}} }
\def\qr{ {Q_{ r}} }

%


\newcommand{\cg}{C_\gamma}
\newcommand{\eg}{E_\gamma}
\newcommand{\ep}{\epsilon}
\newcommand{\g}{\gamma}
\newcommand{\be}{\beta}
\newcommand{\where}{\text{where}}
\newcommand{\for}{\text{for}}
\newcommand{\f}{\frac}
\newcommand\bea{\begin{eqnarray}}
\newcommand\eea{\end{eqnarray}}
\newcommand\p{\partial}
\newcommand\ie{{\emph{i.e.}}}
\newcommand{\bE}{\cx{E}}
\newcommand{\balpha}{\cx{\alpha}}
\newcommand{\angstrom}{\text{\normalfont\AA}}
\newcommand{\braket}[3]{\bra{#1}\;#2\;\ket{#3}}
\newcommand{\projop}[2]{ \ket{#1}\bra{#2}}
\newcommand{\ket}[1]{ |\;#1\;\rangle}
\newcommand{\bra}[1]{ \langle\;#1\;|}
\newcommand{\iprod}[2]{ \langle#1|#2\rangle}
\newcommand{\intl}[2]{\int\limits_{#1}^{#2}}
\newcommand{\logt}[1]{\log_2\left(#1\right)}

\newcommand{\mc}[1]{\mathcal{#1}}
\newcommand{\mf}[1]{\mathbf{#1}}
\newcommand{\mb}[1]{\mathbb{#1}}
\newcommand{\cx}[1]{\tilde{#1}}
\newcommand{\dx}[1]{\hat{#1}}
\newcommand{\blang}{\big \langle}
\newcommand{\brang}{\big \rangle}

\newcommand{\cmnt}[2]{\textbf{\#\#}{\color{#1}#2}\textbf{\#\#}}

\newcommand{\flap}{\mb{L}_{\bar{\kappa}}}
\newcommand{\flapfinitev}{\mb{L}^v}
\newcommand{\flapfinitem}{\mb{L}^p}
\newcommand{\fcurr}{\mb{A}}
\newcommand{\flapFull}{|\Delta|^{3/4}}
\newcommand{\tdir}{f}
\newcommand{\mzeta}{\chi}
\newcommand{\eline}{-----------------------------------------------------------------------------------\\}
\newcommand{\eqa}[1]{\begin{align}#1\end{align}}
\newcommand{\iu}{{i\mkern1mu}}

\begin{abstract}

A many-body interacting system of classical kicked rotor serves as a prototypical model for studying Floquet heating dynamics.    
Having established the fact that this system
 exhibits a long-lived prethermal phase
with quasi-conserved average Hamiltonian
before entering into the chaotic heating regime, 
we use spatio-temporal fluctuation correlation of kinetic energy as a two-point observable
to probe the above dynamic phases. We remarkably find
the diffusive transport of fluctuation in the prethermal regime 
suggesting a novel underlying hydrodynamic picture in a
generalized Gibbs ensemble with a definite temperature that depends on the driving parameter and the initial conditions.  On the other hand,  the heating regime is  characterized by a diffusive growth of kinetic energy 
where the correlation is sharply localized 
around the fluctuation center for all time. Consequently, we attribute non-diffusive and non-localize structure of correlation to the crossover regime, 
connecting the prethermal phase to the heating phase,  
where the kinetic energy displays a complicated growth structure. 
We understand these numerical findings using the notion of 
relative phase matching
where prethermal phase (heating regime)
refers to an effectively coupled (isolated) nature of 
the rotors. \textcolor{black}{We exploit the statistical uncorrelated nature of the angles of the rotors in the heating regime  
to find the analytical form of the correlator that mimics our numerical results in a convincing way.}

\end{abstract}

\maketitle


\textcolor{black}{\emph{Introduction.}}---
In recent years periodically driven isolated systems emerge as an exciting field of research, giving justice to the fact that 
driven systems exhibit intriguing properties as compared to their 
equilibrium counterparts \cite{shirley65,dunlap86,grifoni98}. 
The quantum systems are studied extensively in this context theoretically \cite{goldman14,eckardt17atomic,bukov2015universal} as well as experimentally \cite{Wang453,Experiment2013,Experiment2016,fleury2016floquet,Experiment2017}; for example, dynamical localization \cite{kayanuma08,nag14,nag15}, many-body localization \cite{d13many,d14long,ponte15periodically,ponte15,lazarides15fate,zhang16}, quantum phase transitions \cite{eckardt05superfluid,zenesini09}, Floquet topological insulator \cite{oka09photovoltaic,kitagawa11transport,lindner11floquet,rudner13anomalous,vega19,seshadri19,nag19,nag20a,nag20b}, Floquet topological superconductor \cite{ghosh21a,ghosh21b},
Floquet time crystals \cite{else16floquet,khemani16phase,zhang17observation,yao17discrete}, higher harmonic generation \cite{faisal97,nag17,ikeda18,neufeld2019floquet} are
remarkable nonequilibrium phenomena. Consequently the heating  happens to be very crucial factor as far as  the stability of
the driven systems is concerned \cite{Bilitewski15,Reitter17,Boulier19}.
The consensus so far is that the driven quantum many-body systems heat up to an infinite-temperature state \cite{moessner2017equilibration,luitz17,d14long,Seetharam18} with some exceptions \cite{Prosen98,halder18}. 
However, it has been shown that heating can be suppressed for integrable systems due to infinite number of constants of motion, as manifested through the non-equilibrium steady states \cite{russomanno12periodic,nag14,gritsev2017integrable}. On the other hand,
many-body localized systems prevent heating for their effective local integrals of motion in the presence of interaction and disorder \cite{d13many,d14long,ponte15periodically,ponte15,lazarides15fate}. The high frequency driving is another alternative route to prohibit the heating in the long-lived prethermal region, that grows exponentially with frequency, before heating up at the infinite temperature state
~\cite{choudhury14stability,bukov15prethermal,citro15dynamical,Mori15,chandran16interaction,canovi16stroboscopic,mori2016rigorous,lellouch17parametric,weidinger17floquet,abanin2017effective,else17prethermal,zeng17prethermal,Peronaci18,rajak2018stability,Mori18,Howell19,rajak2019characterizations,abanin2017rigorous}.

Interestingly, the quasistationary  prethermal   state is concomitantly described by an  effective
static Hamiltonian, obtained  using the Floquet-Magnus expansion, in the high-frequency regime \cite{Mori15,canovi16stroboscopic,mori2016rigorous,weidinger17floquet,abanin2017effective,else17prethermal,zeng17prethermal,Peronaci18}. 
Here arises a very relevant question whether the  classical systems exhibit such interesting intermediate prethermal plateau.  Recently, using generic many-body systems of classical chaos theory \cite{rajak2018stability,rajak2019characterizations} and periodically driven classical spin chains \cite{Howell19,Mori18}, the classical systems are also found to demonstrate the  Floquet prethermalization.
Similar to the quantum case, Floquet-Magnus expansion leads to an effective static classical Hamiltonian describing the prethermal phase  where heating is  exponentially suppressed \cite{Howell19,Mori18,rajak2019characterizations,torre2020statistical,Hodson21}.
The prethermal phase is further characterised by 
 generalized Gibbs ensemble (GGE) causing 
hydrodynamic behavior to emerge in the above phase \cite{takato16,mori2018thermalization,deutsch2018eigenstate}.

The framework of fluctuating hydrodynamics 
becomes a convenient tool to investigate the equilibrium transport in classical non-linear systems
\cite{das2014role,Mendl13,Mendl14,spohn2014nonlinear,mendl2015current,kundu16,dhar2019transport,doyon2019generalized,das2020nonlinear}. The integrable (non-integrable) classical systems typically  admit ballistic (non-ballistic) transport \cite{spohn2014nonlinear,das14pre,bastianello2018generalized,doyon18,spohn2020ballistic}. The theory of fluctuating hydrodynamics is also employed to understand the transport in non-linear Fermi-Pasta-Ulam-Tsingou like systems \cite{Mendl13,das14pre}. 
Given the above background, we would like to investigate the non-equilibrium dynamics of fluctuation 
correlation of the kinetic energy
in   a model of interacting classical kicked rotors as a probe to the  hydrodynamic behavior of the problem.
The motivation behind choosing such model is that 
in the limit of large number of particles per site,
a Bose-Hubbard model can be mapped to the above model \cite{rajak2019characterizations}. More importantly, kicked rotor systems can be realized in experiments using Josephson junctions with Bose-Einstein condensates \cite{cataliotti2001josephson}. The time-periodic delta function kicks can be implemented by varying the potential depth and width controlling the intensity of laser light \cite{cheneau2012light,goldman2014periodically}.
 The main questions that we pose in this work are as follows: 
How does a typical fluctuation  
behave in quasi-stationary prethermal states, as well as in the regime where kinetic energy grows in an unbounded
manner \cite{rajak2019characterizations}? Provided the notion of the GGE in the dynamic prethermal regime, does diffusive transport as seen for the case of static Hamiltonian \cite{das2020nonlinear}  persist? Moreover, our questions are  very pertinent experimentally
where Floquet prethermalization has been realized in optical lattice platforms \cite{messer18,rubio20}.

\begin{figure}
	\centering
	\includegraphics[width=\linewidth]{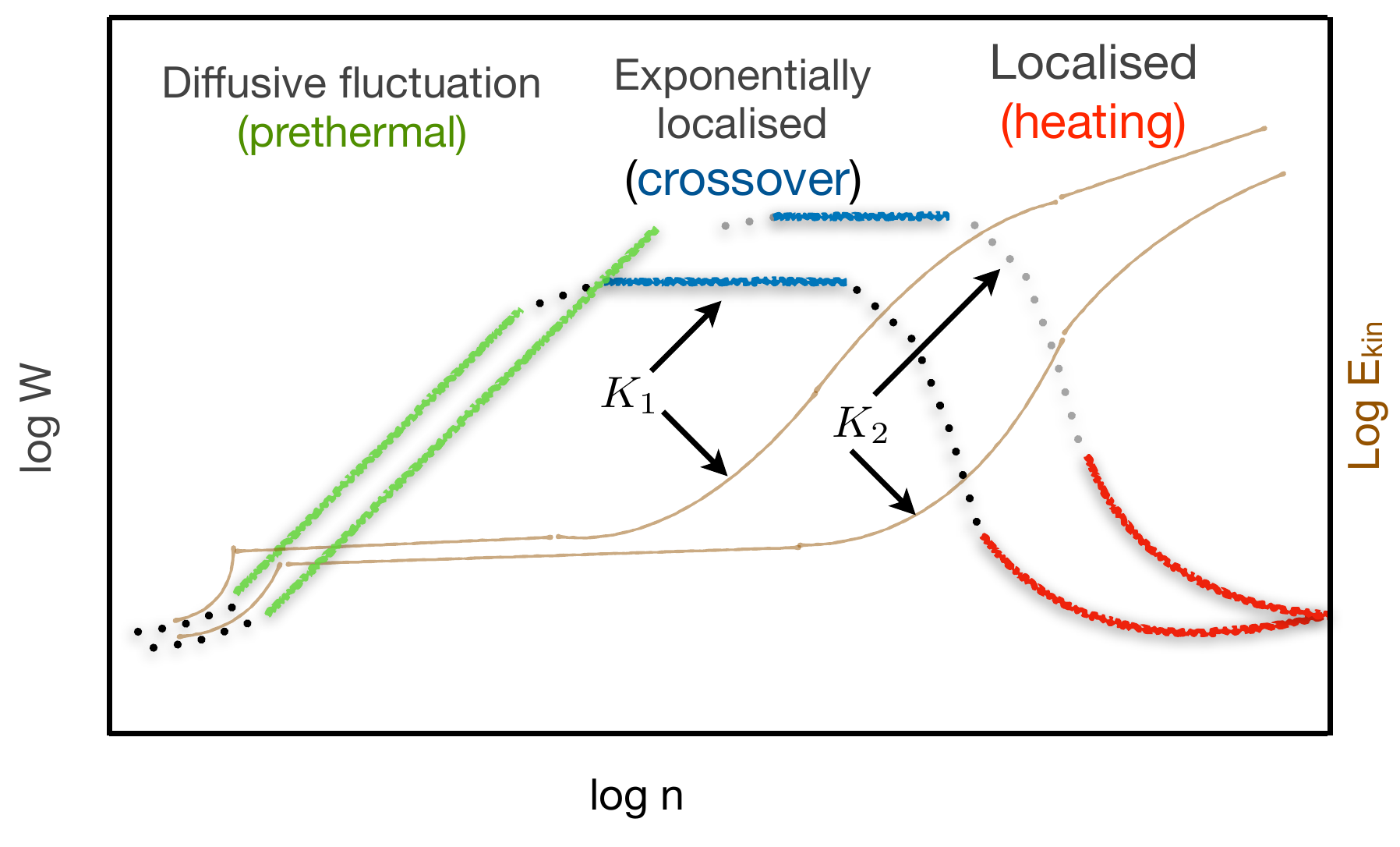}
	\caption{Schematic shows the 
	evolution of relative width of spatio-temporal correlations (\ref{eq_correlation}), indicated by the \textcolor{black}{log of} variance $W$ (left axis),  and the \textcolor{black}{log of} average 
	kinetic energy per rotor $E_{\rm kin}(n)=(1/N)\sum^N_{j=1} \langle p^2_j(n)/2\rangle$ (right axis,  brown solid line),  
	as a function of number of kicks $n$ for two different values of driving parameters $K_1$ and $K_2$. The  kinetic energy is quasi-conserved ($E_{\rm kin}(n) \sim n^0$,  brown solid line) and  the spread of spatio-temporal correlations is self-similar diffusive
	($W \sim n$, green solid line) in the prethermal phase \textcolor{black}{(see Fig.~\ref{fig:PTcorr})}. The intermediate crossover regime is associated with non-diffusive growth of kinetic energy ($E_{\rm kin}(n) \not\sim  n$, brown solid line) and temporally frozen correlations but exponentially localized in 
	space ($W \sim n^{0}$, blue solid line), \textcolor{black}{see Fig.~\ref{fig:IL}}. The diffusive growth of 
	kinetic energy ($E_{\rm kin}(n) \sim  n^1$, brown solid line) corresponds to heating regime where
	 fluctuations tend to be almost $\delta$-correlated in space and time ($W \to 0$, with $n \to \infty$, red solid line), \textcolor{black}{see Fig.~\ref{fig:k1p2}}. Interestingly, the lifetime of prethermal (heating) regime decrease (increases) with  increasing driving parameter $K_1 > K_2$. The different regimes of $W$ are connected 
	 by black (grey) dotted line for $K_1$ ($K_2$). \textcolor{black}{ We note that this plot is a cartoon representation while the exact numerical analysis for the evolution of the kinetic energy is given in Fig.~$2$ of Ref.~\cite{rajak2019characterizations}}.}
	\label{fig:schematic}
\end{figure}
\begin{figure}
	\centering
	\includegraphics[width=\linewidth]{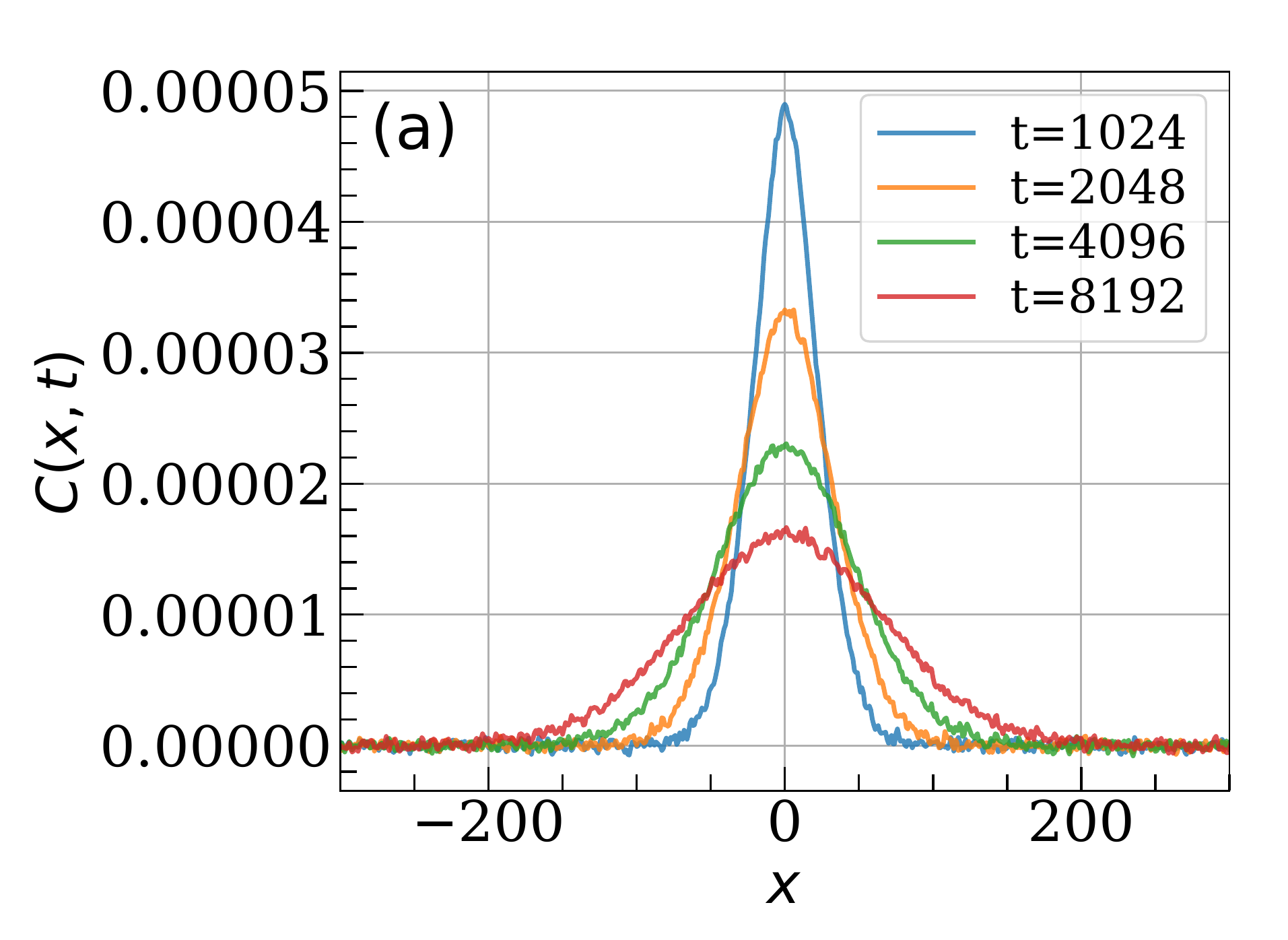}
    \includegraphics[width=\linewidth]{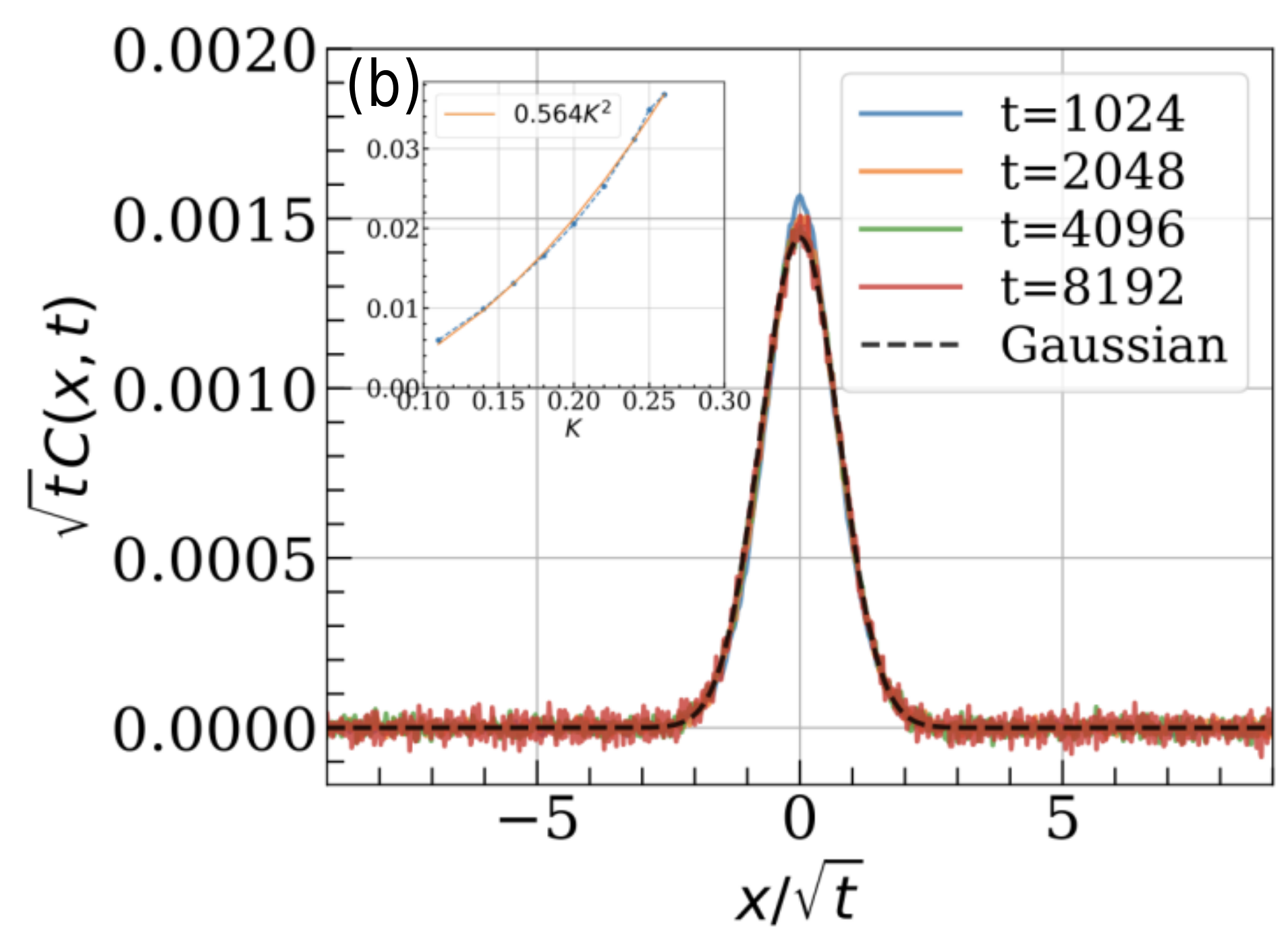}

	\caption{(a) The space-time spreading of kinetic energy correlation (\ref{eq_correlation}) for $K=0.14$ in the prethermal regime.  
	 (b) The diffusive Gaussian scaling of correlation is observed with appropriate renormalization:
	 $C(x,t) =A_K t^{-1/2} f(xt^{-1/2}),~ f(y) = e^{-y^2/2D}/\sqrt{2\pi D}$ with parameters $D\sim 0.727$ and $A_K \sim 0.0026$. Inset shows the quadratic variation of 
	 the total area under the curve $\sum_{x=1}^N  C(x,t)$ with $K$. \textcolor{black}{ The parameters are $N=2048$, and $t_w = 64$}.}
	\label{fig:PTcorr}
\end{figure}
\begin{figure}
	\centering
	
	\includegraphics[width=1.0\linewidth]{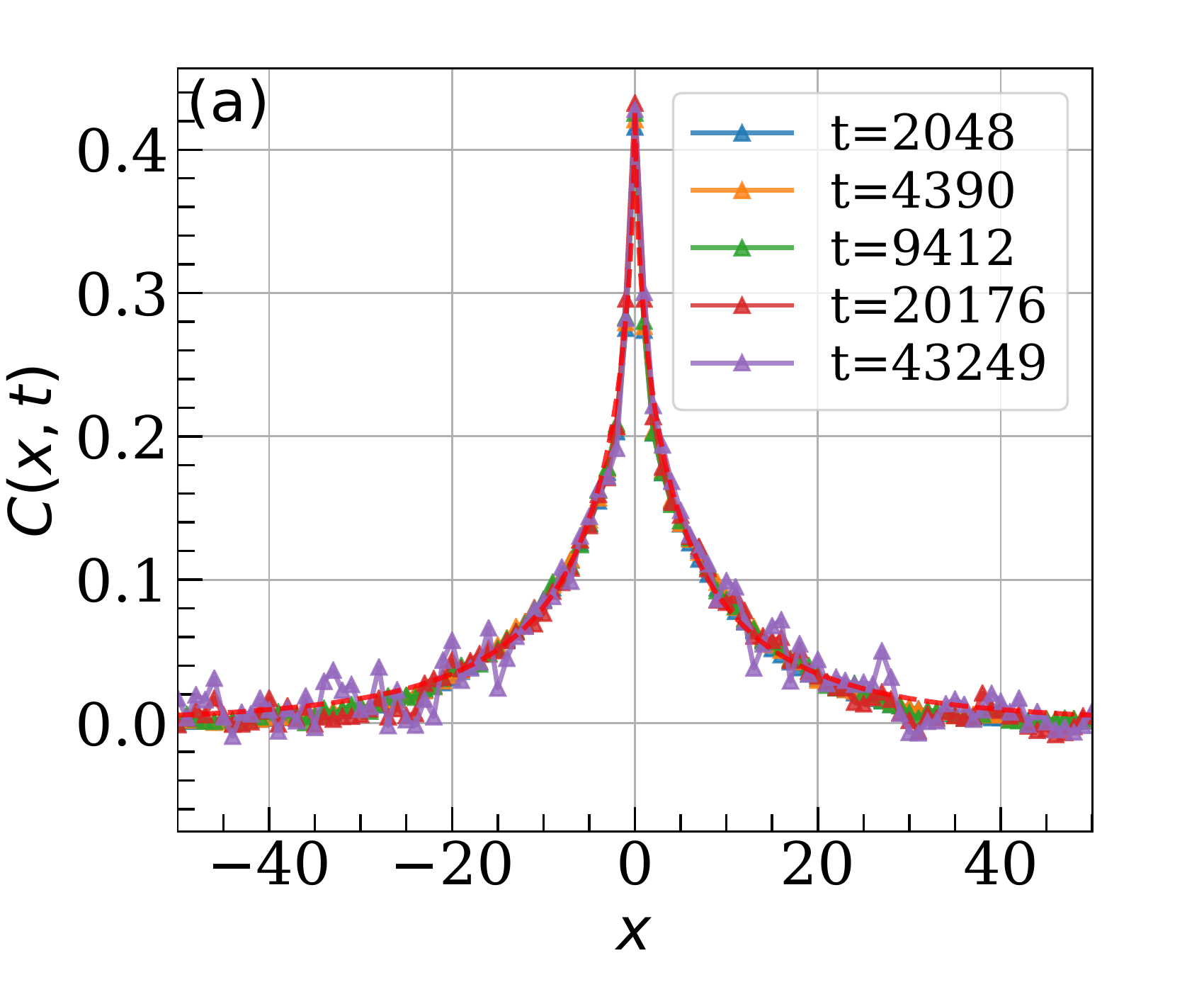}
    \includegraphics[width=1.0\linewidth]{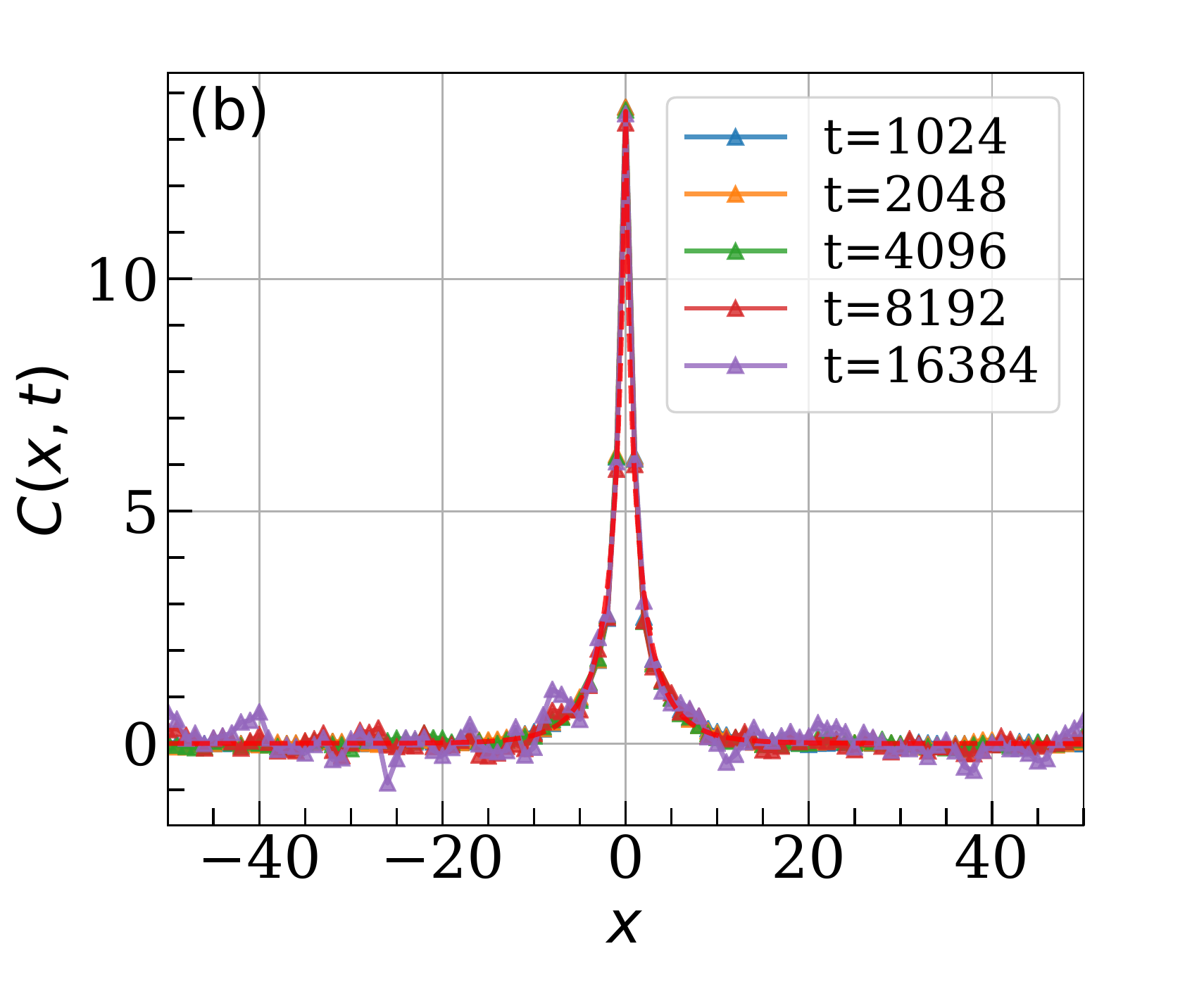}

	\caption{The space-time spreading of kinetic energy correlation (\ref{eq_correlation}) for  (a)  $K=0.45$, \textcolor{black}{$t_w = 142$} and  
	$K=0.7$,  \textcolor{black}{$t_w = 64$} (b), respectively, in the intermediate crossover regime. The correlation distribution is fitted with stretched exponential $C(x,t) \sim \frac{A_K}{2 \alpha ^{-1/\beta } \Gamma \left(1+\frac{1}{\beta }\right)} e^{-\alpha |x|^\beta}$ with  $\alpha \sim 0.36~[1.05], \beta \sim 0.63~[0.65], A_K \sim 5.16~[41.51]$ for (a) [(b)]. \textcolor{black}{The system size is $N=2048$ for both cases.}}
	\label{fig:IL}
\end{figure}
\begin{figure}
	\centering
	\includegraphics[width=\linewidth]{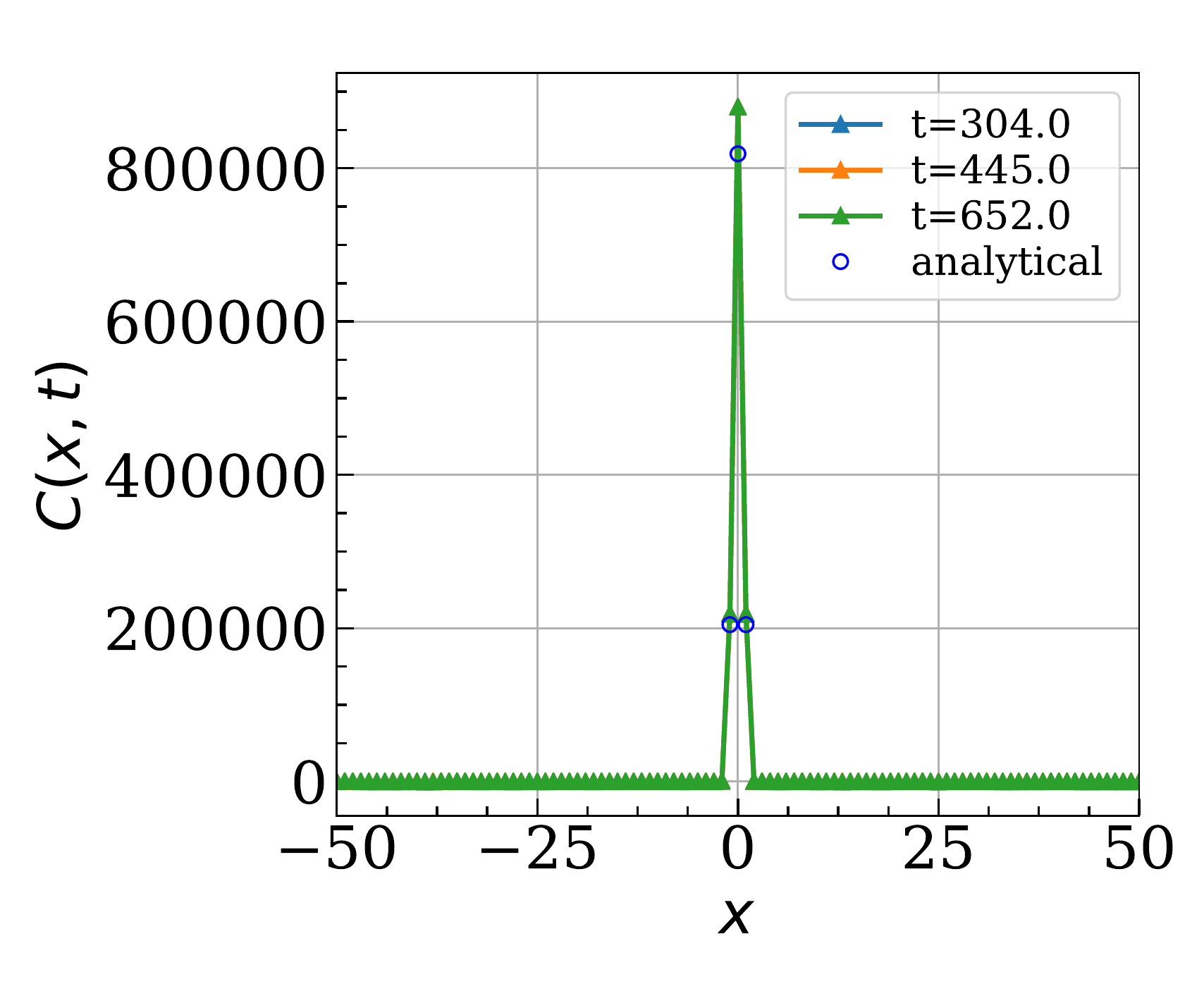}
	\caption{The correlations (\ref{eq_correlation}) become
		fully localized such as nearly a $\delta$-function in the heating region for $K=3$, \textcolor{black}{with  $t_w=142$, $N=2048$. The blue circles denote the correlation values as a function of initial wait time and the kick strength given by Eq.~\ref{heating_corr}. } }
	\label{fig:k1p2}
\end{figure}

Given the fact that the time-dynamics of the kinetic energy for the above system can be divided into three different temporal regimes depending on the nature of its growth \cite{rajak2018stability,rajak2019characterizations}, in this work, while numerically investigating the propagation of fluctuation (\ref{eq_correlation}) through the system as a function of time,  
we show that these dynamical regimes are characterized by distinct space-time behavior of the kinetic energy fluctuation correlation (see Fig.~\ref{fig:schematic}).  
Following the initial transient, the system enters into the prethermal regime, characterized by almost constant kinetic energy with exponentially suppressed heating, 
where  the fluctuation spreads over space diffusively as a function of time (see Fig.~\ref{fig:PTcorr}).  The spatio-temporal correlation becomes 
Gaussian  whose  variance $W$  increases linearly with time. 
Once the system starts absorbing energy from the drive, the
fluctuation becomes exponentially localized around the site of disturbance and 
temporally frozen referring to the constant nature of $W$ with time (see Fig.~\ref{fig:IL}). We refer 
this intermediate window as a crossover region that connects the spatially and temporally  quasi-localized behavior of correlations at long time (see Fig.~\ref{fig:k1p2}) with the  
prethermal phase. In that quasi-localized phase, $W$ decays to vanishingly small values while  
the kinetic energy of 
the system grows linearly with time. Therefore, the  kinetic energy localization (diffusion) corresponds to the diffusion (localization) of fluctuation correlation.  We qualitatively understand the underlying energy  absorption mechanism in prethermal phase
based on the hydrodynamic description. 
Our study considering fluctuation correlation of kinetic energy as a two-point observable, reveals new insight to the dynamic phases that are not accounted by the one-point observables. Moreover, 
dynamic features such as  quasi-localization and localization of spatio-temporal correlations  do not have any static analogue.


\textcolor{black}{\emph{Model and correlation fluctuation.}}---
We consider a generic non-equilibrium many-body system of classical chaos theory as given by
 \cite{kaneko89diffusion,konishi90diffusion,falcioni91ergodic,chirikov1993theory,chirikov97arnold,mulansky11strong,rajak2020stability}.
\eqa{H&=\sum_{j=1}^{N}\left[\frac{p_{j}^{2}}{2} -\kappa \sum_{n=-\infty}^{+\infty} \delta(t-n \tau) V\left(r_j\right) \right]
,\label{eq_hamiltonian}}
where stretched variable $r_j = \phi_{j}-\phi_{j+1}$ and $V(r_j) = 1+\cos(r_j)$. Here $\phi_j$, $j=1,\cdots,N$, are the angles of the rotors 
and $p_j$ are the corresponding angular momenta. The parameter $\kappa$ denotes the interaction as well as kick strength, 
and $\tau$ is the time period of delta kicks. The system described in Eq.~(\ref{eq_hamiltonian}) can have infinite energy 
density due to the unbounded nature of kinetic energy. 
We note that the total angular momentum of the 
system is an exact constant of motion, \textcolor{black}{because the Hamiltonian in Eq.~(\ref{eq_hamiltonian}) is invariant under a global translation $\phi_j\rightarrow\phi_j+\alpha$, $\alpha$ being an arbitrary real number. Moreover, the Hamiltonian has discrete time translation symmetry $H(t)=H(t+\tau)$.}
Using classical Hamilton's equations of motion, one can get the discrete maps of $\phi_j$ and $p_j$ between $n$-th and $(n+1)$-th kicks:
\eqa{
	p_j(n+1) &= p_j(n) - \kappa \left( V'(r_{j-1}) -V'(r_j) \right)\nonumber\\
	\phi_j(n+1) &= \phi_j(n) +  p_j(n+1) \tau.
	\label{eq_motion}
}
Here $V'$ describes derivative of $V$ with respect to $r_j$
evaluated after $n$-th kick.
 We consider periodic boundary conditions $\phi_{N+i} = \phi_i$. 
From Eq.~(\ref{eq_motion}), it can be noticed that the dynamics of the system is determined by only one dimenisonless 
parameter, $K=\kappa\tau$ that we use for all our further calculations ~\cite{rajak2018stability,rajak2019characterizations}.

We compute here the spatio-temporal correlation of kinetic energy fluctuations, defined by
\eqa{ C(i,j,t,t_w) = \frac{1}{4} \left[  \la{p_i^2(t)}{p_j^2(t_w)} \ra - \la {p_i^2(t)} \ra \la {p_j^2(t_w)} \ra \right],
\label{eq_correlation}}
where $i$ and $j$, respectively, represent the positions of $i$-th and $j$-th rotors; $t$ ($t_w$) represents an arbitrary final time (initial waiting time). \textcolor{black}{We always consider $t>t_w$ throughout the paper.}
The symbol $\langle..\rangle$ denotes the average over the initial conditions where $\phi_j(0)$ are chosen from a uniform distribution $\in [-\pi,\pi ]$, and
the corresponding momenta, $p_j(0)=0$ for $j=1,\cdots,N$. 
The spatio-temporal correlation captures how a typical small perturbation \textcolor{black}{applied at time $t_w$}
spreads in space $x \equiv i-j$ (with translation symmetry)
and time $t$ through the system. \textcolor{black}{
We refer to the correlator in
Eq.~(\ref{eq_correlation}) as $C(x,t)$ while investigating below.
The system shows an exponentially long prethermal state where the kinetic energy 
becomes almost constant, and eventually heats up after a crossover regime when the kinetic energy grows linearly with time~\cite{rajak2018stability,rajak2019characterizations}.
We have further analyzed different temporal regimes investigating the behavior of spatio-temporal correlation.
The prethermal state can be characterized appropriately by a time-averaged Hamiltonian (see Eq.~(\ref{eq_eff_ham})). 
Therefore, although, the driven system breaks continuous time-translation symmetry, it preserves an effective
time-translation symmetry inside the prethermal regime due to quasi-conservation of Floquet Hamiltonian at high frequencies. Thus, the spatio-temporal correlator becomes a function 
of space $x \equiv i-j$ and time $t \equiv t-t_w$ inside the prethermal regime. However, for other two regimes where the 
total energy is not a constant of motion, the correlator generally becomes a function of both $t$ and $t_w$ in addition to $x$.
}

\textcolor{black}{We have summarized our main result of spatio-temporal correlation and its connection with the evolution of kinetic 
energy schematically in Fig.~\ref{fig:schematic}. In this context, we consider the width of the spatio-temporal correlations, i.e., 
variance \eqa{W =\frac{\sum_{x=-N/2}^{N/2}  x^2 C(x,t)}{\sum_{x=-N/2}^{N/2}  C(x,t)},} to characterize different dynamical regimes.
In order to  measure the relative width of the fluctuations, the  appropriate normalization of the distribution as described by the denominator is  crucial. In the prethermal regime, the denominator is independent of time due to conservation of energy, while in the other regimes, the denominator is time dependent and normalizes the distribution at all times.
We have schematically drawn  the evolution of $W$ in Fig.~\ref{fig:schematic} by acquiring detailed knowledge about spatio-temporal correlation
in different phases as discussed below.
}

The observables are averaged over $10^{5}$ to $10^{6}$  initial conditions. Otherwise specifically mentioned 
in our simulations, we fix $\tau=1$ and $N=2048$. This makes $t=n$, and we use these terms interchangeably. 
The lifetime of the prethermal state for such systems (see Eq.~(\ref{eq_hamiltonian})) increases exponentially in $1/K$ \cite{rajak2019characterizations}. 
For small values of $K$, the prethermal state persists for astronomically large time, thus making the numerical calculation 
extremely costly to probe all the dynamical phases by varying time.  In order to circumvent this problem, we choose to tune $K$ such that  
the lifetime of the prethermal state can be substantially minimized and we can investigate the  phase where fluctuations get localized  within our numerical facilities. 
However, provided the distinct nature of these regimes, our findings would remain unaltered if one addresses them by varying time only. 
In our numerical calculations, we choose both $t,~t_w$  within the same phase.

\textcolor{black}{ We would like to emphasize the choice of $t$ and $t_w$ such that the fluctuation correlation of kinetic energy can behave distinctly in different regimes. 
There can be some 
quantitive but no qualitative changes in the correlator for $t$ and $t_w$ chosen from same regimes while quantitive changes are observed for $t$ and $t_w$ chosen from different regimes. In order to give an idea about the choice of $t$ and $t_w$, we exemplify a situation with $K=0.3$ where  $10^1<t,~t_w<10^3$ for prethermal regime,  $6\times10^3<t,~t_w<10^4$ for the crossover regime  and $2\times10^5<t,~t_w<10^7$  for the heating regime
\cite{rajak2019characterizations}. As discussed above the temporal width of various regimes  
vary with $K$ and hence $t$ and $t_w$ are needed to be appropriately chosen within the same regimes. 
 }

\textcolor{black}{\emph{Results.}}---
We first focus on the spreading of fluctuation  (\ref{eq_correlation}) in
the prethermal phase that is denoted by the green solid lines in Fig. \ref{fig:schematic}. 
The prethermal phase can be described by a 
GGE with the total energy as a quasi-conserved quantity
\cite{rajak2019characterizations}. 
In terms of the inverse frequency Floquet-Magnus expansion, the lowest order term of the Floquet Hamiltonian is 
the average Hamiltonian that governs the prethermal state at high frequency, given by 
\eqa{H^{*}=\frac{1}{\tau} \int_{0}^{\tau} H(t) d t =\sum_{j=1}^{N}\left[\frac{p_{j}^{2}}{2}-\frac{\kappa}{\tau} \left(1+\cos \left(\phi_{j}-\phi_{j+1}\right)\right)\right].
\label{eq_eff_ham}}
Employing the notion of GGE, the 
composite probability distributions   can be written as
\eqa{ P^*(\{p_j,r_j\}) =\frac{1}{Z^*}  \prod_{j=1}^{N} e^{- \big(\frac{p_j^2}{2} - \frac{\kappa V(r_j)}{\tau }\big)/T^* },
 \label{eq_distribution}}
where $Z^*$ is the partition function for the GGE and 
$T^*$ is the temperature associated with prethermal phase. 
Given the particular choice of the initial conditions here, the prethermal temperature is found to be $T^* = 0.938363 \frac{K}{\tau^2}$ \cite{rajak2019characterizations}; for more details see Appendix \ref{GGE_appendix}.
Moreover, this description of the GGE does not depend on the number of rotors $N$, thus indicating the thermodynamic stability of this phase.

We associate the prethermal phase with the  diffusive spatio-temporal spread of kinetic energy correlation as shown in 
Fig.~\ref{fig:PTcorr} (a). A relevant renormalization   
of  $x$ and $y$-axes with time yields the following 
scaling form of the correlation:
 $C(x,t) \sim A_K t^{-1/2} f\big(xt^{-1/2} \big)$, 
with $f(y) = e^{-y^2/2D}/\sqrt{2\pi D}$, as depicted in  Fig.~\ref{fig:PTcorr} (b); for more \textcolor{black}{detailed discussions}  see Appendix \ref{diffusion_appendix}. We thus find  that the correlation at different space time collapse together.
Here,  $A_K$  denotes the amplitude of  the Gaussian distribution respectively for a given value of $K$. The diffusion constant $D$ is a measure of variance $W$, being weakly dependent on the parameter $K$, grows linearly 
with time $W \sim t$.  
One can observe that the fluctuation spreads 
in a way such that the area under the  correlation curves keep their area constant i.e. the sum rule $\sum_x C(x,t)$ is approximately  independent of  time in the prethermal phase. 
The sum rule determines $A_K = \sum_{x=-N/2}^{N/2}  C(x,t) \approx \sum_{x=-N/2}^{N/2}  C(x,0) = \delta_{x,0}\left(\la p^4 \ra - \la p^2 \ra^2 \right)/4= 0.4402K^2/\tau^4 \delta_{x,0}$, by considering the fact that the energy absorption is exponentially suppressed in the prethermal regime \cite{rajak2019characterizations}; for more details see Appendix \ref{GGE_appendix}. This supports our numerical result of quadratic growth of $A_K$ as shown in the inset of 
Fig.~\ref{fig:PTcorr} (b). The apparent  $27\%$ mismatch
in the  prefactor of $A_K$ with the numerically value might be due to the fact that  exponentially slow variation of the kinetic energy in the prethermal phase is not taken into account theoretically.

The energy correlations of static rotor system at high temperature platform exhibit diffusive transport \cite{lepri2016thermal,dhar2008heat}. This is in resemblance with the present case of Floquet prethermalization at high frequency. Owing to the quasi-validity of equipartition theorem in the GGE picture \cite{rajak2018stability,rajak2019characterizations}, 
the kinetic and potential energy behave in an identical fashion. As a result, the correlation of total energy qualitatively follows the correlation of kinetic energy in the prethermal regime. 

We now investigate the spatio-temporal evolution of correlation (\ref{eq_correlation}) in the intermediate crossover  regime, designated by  the blue solid line in Fig. \ref{fig:schematic}, that lies 
between the prethermal and the heating region  of 
kinetic energy.  The system starts to absorb energy from the drive through many-body resonance channels causing the 
kinetic energy to grow in sub-diffusive followed by super-diffusive manner \cite{rajak2018stability,rajak2019characterizations}.
However, the probability of the occurrence of such resonances decreases exponentially with $1/K$ in the high frequency limit. 
We find  that the oscillators are maximally correlated with each other at $x=0$ and falls rapidly to zero in two sides $x$ as shown in  Fig.~\ref{fig:IL} (a) and (b), for $K=0.45$ and $0.7$, respectively. To be precise, correlation  
 decays  stretched exponentially in  short distances: 
 $C(x,t) \sim A_K e^{-\alpha |x|^\beta}$ while it falls  exponentially (i.e., more rapidly than stretched exponential) in  long distances: $C(x,t)  \sim e^{-\gamma |x|} $; for more details see Appendix \ref{crossover_appendix}. Here, $\beta$ and $\gamma$ weakly depend on $K$ referring to the fact that driving parameter $K$ can in general  control the spatial spread of fluctuation.  
These profiles do not change with time within the crossover region
referring to the fact that variance of the spatial correlation distribution remains constant with time $W \sim t^0 $.  This allows 
 us to differentiate it from the diffusive transport that occurs in the prethermal phase.
However, with increasing time, one can observe that long distance correlation becomes more noisy
leaving the spatial structures qualitatively unaltered.


At the end, we discuss the time zone where the average kinetic energy shows unbounded chaotic diffusion, as denoted by  the red solid line in Fig.~\ref{fig:schematic}, resulting in 
the effective temperature to increase linearly with time~\cite{rajak2018stability, rajak2019characterizations}. The correlation of the kinetic energy is fully localized in space and temporally frozen as shown in Fig.~\ref{fig:k1p2}). To be precise, the correlation 
is nearly a $\delta$-function centered around $x=0$ i.e., fluctuation gets localized at the site of disturbance for all time. In this regime, the system shows fully chaotic behavior in the phase space and the angles of the rotors
become statistically uncorrelated both in space and time.
It is noteworthy that 
there is no description of average Hamilton exist here as  the inverse frequency Floquet-Magnus expansion does not converge \cite{Mori18,Howell19}. 
In contrast to the prethermal phase, the amplitude  of correlation peak in the crossover and heating regime increases  as $K^\eta$ with $\eta>2$. On the other hand, the variance in the heating regime becomes decreasing function of $n$, precisely, $W \to 0$ for $n \to  \infty$ that is markedly different from the behavior of $W$ in remaining two 
earlier  regimes. Finally, we stress that our findings in this heating regime do not suffer from finite size effect suggesting the 
thermodynamic stability of this phase.

\textcolor{black}{It is noteworthy  that $\la{p_i^2(t)}{p_j^2(t_w)} \ra$ and $\la{p_i^2(t)}\ra  \la{p_j^2(t_w)} \ra$  in the kinetic energy fluctuation both individually depend on time $t$ and the initial waiting time $t_w$. On the contrary, the connected part i.e., $C(i,j,t,t_w)$ as a whole does not depend on $t$ instead depends on $t_w$ such that the peak height of $C(i,j,t,t_w) $-profile increases with $t_w$. This can be physically understood as an initial value problem in terms of $p_{i,j}(t_w)$ for the rotors that subsequently uncorrelated in the heating region.   
The peak value of the correlator is also found to be dependent on the coupling parameter $K$.  
In this region, the time-independent
correlator effectively freezes into three discrete spatial points i.e., at the site of disturbance with $i=j$ and the remaining two adjacent sites with $i=j\pm 1$.}

\textcolor{black}{We shall now exploit the assumption of statistically uncorrelated (both in space and time) nature of the rotors to shed light on this intriguing behavior analytically. In terms of the stretched variable $r_j(n)=\phi_{j+1}(n)-\phi_j(n)$, the assumption leads to the following mathematical form 
\begin{eqnarray}
&&\langle\sin(r_i(n))\sin(r_j(m))\rangle=\frac{1}{2}~\delta_{n,m}\delta_{i,j},\nonumber \\
&&\langle\sin(r_i(n))\sin(r_j(n') \sin(r_k(m))\sin(r_l(m'))\rangle=\nonumber \\
&&\frac{1}{4}~(\delta_{n,n'} \delta_{m,m'} \delta_{i,j} \delta_{k,l} + \delta_{n,m} \delta_{n',m'} \delta_{i,k} \delta_{j,l} \nonumber \\
&&+ \delta_{n,m'} \delta_{n',m} \delta_{i,l} \delta_{j,k})
+ \frac{3}{8} \delta_{n,n'} \delta_{m,m'} \delta_{m,n} \delta_{i,j} \delta_{k,l} \delta_{i,k}
\label{approx}
\end{eqnarray}
where the average is carried over different initial conditions, $i,j,k,l$ represent the position of the rotor and $n,n',m,m'$ denote the various times $t$'s in terms of the number of  kicks. 
We note that $r_j(n)$ becomes uniform random variable in the heating region. 
The momentum of $i$-th rotor at any time 
$t$ can be formulated from the equation of motion (Eq.~\ref{eq_motion}). 
The first term of the  kinetic energy fluctuation correlator with $t>t_w\gg1$ can be calculated in the heating regime as follows \begin{align}
&\langle p_i^2(t)p_j^2(t_w)\rangle 
=K^4\Big(2~ t_w^2 \delta_{i,j} +  \frac{1}{2}~ t_w^2 \delta_{i,j+1} + \frac{1}{2} t_w^2 \delta_{i,j-1} \nonumber \\
& + \frac{3}{4} t_w  \delta_{i,j} + \frac{3}{8}~ t_w \delta_{i,j+1} + \frac{3}{8}~ t_w \delta_{i,j-1}+
tt_w \delta_{i,i} \delta_{j,j} 
\Big).
\label{heating_term1}
\end{align}
The last term reminds us the diffusive growth of kinetic energy $\langle p_i^2(t)\rangle=K^2 t$ that is obtained in the heating region. Therefore, the kinetic energy fluctuation correlation (Eq.~(\ref{eq_correlation})) takes the following form 
\begin{align}
&C(i,j,t,t_w)
=K^4\Big( \frac{t_w^2}{2} \delta_{i,j} +  \frac{3t_w}{16}  \delta_{i,j}+
\frac{t_w^2}{8} \delta_{i,j+1} +  \frac{3t_w}{32} \delta_{i,j+1}\nonumber \\
&+ \frac{t_w^2}{8} \delta_{i,j-1} +  \frac{3t_w}{32} \delta_{i,j-1}\Big )\nonumber \\
&= K^4\Big( \frac{t_w^2}{2} \delta_{i,j} + 
\frac{t_w^2}{8} \delta_{i,j+1} + \frac{t_w^2}{8} \delta_{i,j-1} \Big ) + O(t_w)
\label{heating_corr}
\end{align}
}

\textcolor{black}{We note that for $t_w\gg 1$, the $O(t^2_w)$ terms dominate over $O(t_w)$ term. 
This clearly suggests that the assumption of uncorrelated nature of rotors is able to mimic the numerical outcome  convincingly in the heating region i.e., $C(x,t)\ne 0$ for $x=0$ and $\pm 1$. In Fig.~\ref{fig:k1p2}, we compare the numerical results with the prediction of Eq.~\ref{heating_corr} and find a match within $6\%$ error. 
The above result is derived considering the assumption that rotors are always uncorrelated under driving. 
As a result, the independent rotor approximation works better to explain the numerical outcomes
for higher values of $K$ as the driven system enters into the heating region quite early.
Most importantly, the analytical form in Eq.~(\ref{heating_corr}) correctly captures the value of the correlator at the adjacent sites of the disturbance $i=j\pm 1$
drops to $\frac{1}{4}$ of the peak value at $i=j$, as observed in Fig.~\ref{fig:k1p2}. 
The non-zero value of correlation in the adjacent sites might be the effect of nearest-neighbor interactions of the system. 
In addition, the correlator is independent of final time $t$, whereas it depends on $t_w$, when the disturbance is applied on the system. 
It indicates that the correlator depends on the value 
of momentum at $t=t_w$, from where we start measuring the correlator, but correlator does not spread further with time since the rotors are effectively uncorrelated in this regime.
The detailed analytical derivation is presented in Appendix \ref{heating_analysis}. 
}



\textcolor{black}{We here stress that Fig.~\ref{fig:schematic}, being  a cartoon representation, can compactly demonstrate the results shown in Figs.~\ref{fig:PTcorr}, \ref{fig:IL}, and \ref{fig:k1p2} as well as shed light on the physical understanding. We superimpose 
the time evolution of the variance $W$ of the correlation $C(x,t)$, obtained from analyzing the correlation profile in Figs.~\ref{fig:PTcorr}, \ref{fig:IL}, and \ref{fig:k1p2}, with the kinetic energy $E_{\rm kin}(n)$ that has already been reported \cite{rajak2019characterizations}. In the prethermal regime, $E_{\rm kin}(n) \sim n^0$ and the self-similar Gaussian scaling of $C(x,t)\sim f( x t^{-1/2})$ suggests $ W \sim n $ (see Fig.~\ref{fig:PTcorr}(b)). In the crossover region, $E_{\rm kin} \not \sim n$ and the temporally frozen stretched exponential  
$C(x,t)\sim f(x,t^0)$ yields $W \sim n^0$ (see Fig.~\ref{fig:IL}). In the heating region, $E_{\rm kin} \sim n$, fluctuations become 
nearly $\delta$-correlated but frozen in time $C(x,t)\sim f(x,t^0)\delta_{x,l} $ with $l \in [0,\pm 1]$, leads to the following situation $W \to 0$ as $n\to \infty$ (see Fig.~\ref{fig:k1p2}). 
Therefore, the disjoint nature between the spatial and temporal structures of the correlation in crossover and heating regions causes distinct features in $W$ as compared to its combined spatio-temporal nature for prethermal region. 
}

\textcolor{black}{\emph{Discussions.}}---
This is clearly noted in our study that the spatio-temporal correlation provides
a deep insight to characterize different dynamical phases.
The phase matching between adjacent rotors, captured by stretched variables $r_j(n)=\phi_{j+1}(n)-\phi_j(n)$,  play very important 
role in determining the nature of spreading of the fluctuations.
The time-evolution
of the stretched variable, following the equations of motion (\ref{eq_motion}), is given as $r_j(n+1)=r_j(n)+(p_{j+1}(n+1)-p_j(n+1))\tau$. Upon satisfying the resonance condition $p_j-p_{j+1}=2 \pi m/\tau$ with $m$ as an integer number, 
the stretched variable rotates by $2\pi$ angle
between two subsequent kicks. When all the rotors go through these resonances, their relative phase matching is lost and the eventually the coupled rotor system turned into an array 
of uncoupled independent rotor. At this stage, 
the system absorbs energy from the drive at a constant rate
in an indefinite manner. This is precisely the case for heating up regime where fluctuation correlations do not spread in time and space. On the other hand, the resonances are considered to be extremely rare events 
in the prethermal regime suggesting the fact that 
relative phase matching between 
adjacent rotors allows the fluctuation to propagate diffusively throughout the system in time.
The notion of the time independent average Hamilton in the prethermal region might be related to the fact that 
all the rotors rotate with a common collective phase and eventually controlled by an underlying synchronization phenomena  \cite{nag17,khasseh19}.

Coming to the {phenomenological} mathematical description, 
the exponential suppressed heating and the validity of the constant sum-rule in the prethermal phase  suggest a hydrodynamic diffusion picture (with diffusion constant $D$) for the fluctuation \cite{spohn14,spohnKPZ,dhar2021revisiting,Norman20}:
\eqa{ \p_t u(x,t) = \p_x \bigg( \frac{D}{2}~ \p_x u(x,t)
	+ B  ~ \zeta (x,t) \bigg )
	\label{eq_diffusion}}
 where $u(x,t) = \frac{1}{2} \left(p^2(x,t) - \la p^2(x,t) \ra\right)$ such that $\la p^2(x,t) \ra \neq \la p^2(x, t') \ra$. The  conservative noise $\zeta(x,t)$ of strength $B$
 is delta correlated in space and time $\la \zeta(x,t) \zeta(x',t')\ra =  \delta_{xx'} \delta(t-t')$. 
In the high-frequency limit, the  equilibrium fluctuation dissipation relation can be extended to long-lived prethermal regime: $B^2 \sim D { T^*}^2$; for more details see Appendix \ref{hydrodynamic_appendix}.  Given the plausible assumption that the  noise part $B$ of the fluctuating current increases with increasing $K$, one can understand the diffusion process in a phenomenological way. The diffusion constant $D$ is then considered to be independent of $K$ while prethermal temperature is determined by $K$, as observed in GGE picture.
Moreover, the self-similar Gaussian nature of fluctuation correlation in the prethermal phase can be understood from the solution of Eq.~\ref{eq_diffusion} such that $ C(x,t) \sim  \la u^2(0,\tau_i) \ra e^{-x^2/(2Dt)} /\sqrt{2\pi D t}$. In the other limit inside the 
infinite temperature heating regime, the  non-diffusive transport leaves the $u(x,t)$ to be $\delta$ correlated in space while almost frozen in time. Now there is an extended  crossover region, connecting the prethermal state to heating regime, where the diffusion equation does not take such simple form causing the system to exhibit an amalgamated
 behavior.

\textcolor{black}{Before we conclude, we would like to re-emphasize that unlike the equilibrium case $t_w$ plays crucial role for the non-equilibrium case. To be more precise, only relevant time variable is the relative time $t \equiv t-t_w$ for equilibrium case respecting time translation symmetry. This is not true in the present case and the correlation is expected to exhibit complex structure for arbitrarily chosen $t$ and $t_w$ with $t>t_w$.
Thanks to the GGE description of the prethermal region, we can consider  the relative time $t \equiv t-t_w$ as the appropriate variable.
One can think of the average Hamiltonian embeds an effective time translation symmetry in the prethermal phase.  
However, the effective GGE description fails for crossover and heating region, resulting in the fact that the  two time instants  $t$
and $t_w$ are equally important there.
The effective time translation symmetry is no longer valid in the above two regions.
This is what we clearly see for the heating regime where the peak value of the correlator depends on $t_w$.  }

\textcolor{black}{\emph{Conclusions.}}---
In conclusion, our study demonstrates that a typical two-point observable such as,
the fluctuation correlation of kinetic energy, can be scrutinized to probe different dynamic phases of classical many-body kicked rotor system \cite{computation}. It is indeed counter intuitive that the average kinetic energy per rotor and their spatio-temporal correlations, being derived from the former quantity, yet yield opposite  behavior. 
The GGE description of prethermal phase obeys diffusive transport where spatio-temporal correlation follows
self similar Gaussian profile.  During this diffusion, the correlation curves keep their area constant
due to the quasi-conservation of the kinetic energy. In the long time limit where kinetic energy grows diffusively, fluctuation interestingly becomes frozen in space and time. 
\textcolor{black}{We have also  calculated the behavior of  kinetic energy fluctuation analytically using the statistically uncorrelated nature of the angles of the rotors inside the heating regime and find a good match with the numerical results.}
There exists an extended crossover region where kinetic energy increases in a  complicated way exhibiting both sub-diffusive to super-diffusive nature. The fluctuation interestingly shows a rapidly (slowly) decaying short (long) range stretched (regular) exponential localization. In this case the  fluctuation  does not have any time dynamics. Therefore, correlated phenomena in prethermal phase gradually assembles to completely uncorrelated heat up phase through a crossover region. These non-trivial phases of matter are the consequences of the driving and do not have any static analogue. Provided the understanding on long range quantum systems  \cite{Lerose19,Saha19,Tomasi19,garg19,modak20a,modak20b}, it would be interesting to study the fate of the above phases along with the crossover region in long-range classical systems.
\textcolor{black}{Furthermore, various intriguing nature of correlators can be observed for
$t$ and $t_w$ chosen from different phases that is beyond the scope of the present study.}
The microscopic understanding of hydrodynamic picture and fluctuation-dissipation relation in dynamic systems are yet to be  extensively analyzed in future.


{\it{Acknowledgement} }:
 We thank Andrea Gambassi and Emanule Dalla Torre for reading the manuscript and their constructive  comments. A.R. acknowledges UGC, India for start-up research grant F. 30-509/2020(BSR). AK would like to thank the  computational facilities Mario at ICTS and Ulysses at SISSA.

\appendix

\section{Gibbs distribution in prethermal phase}
\label{GGE_appendix}

Assuming that in the prethermal region, the ensemble is goverened  by a Gibbs ensemble with
\eqa{ P^*  = (Z^*)^{-1} e^{-\left(H^* + \gamma \bar{p} + \delta \cx{r} \right)/\left( T^{*}\right)}, }
where $\bar{p}=\sum_{j=1}^Np_j$ and $Z^* = \int dp dr e^{-\left(H^* + \gamma \cx{p} + \delta \cx{r} \right)/T^*}$ are the total angular momentum and 
the GGE partition function of the system, respectively~\cite{rajak2019characterizations}.
Equating the initial energy   with the GGE average energy, we obtain the effective temperature of the system.  The kinetic energy $E_{\mathrm{kin}}$ of the driven system in the prethermal phase is given by 
\eqa{E_{\mathrm{kin}}=\frac{1}{N} \sum_{j=1}^{N} \int \frac{p_{j}^{2}}{2 Z^*} \prod_{j^{\prime}=1}^{N} e^{-\left(p_{j^{\prime}}-\cx{p}\right)^{2} /\left(2 T^{*}\right)} d p_{j^{\prime}}=\frac{T^{*}+\tilde{p}^{2}}{2},}
where $\cx{p}=\gamma T^*/N$ with $\gamma$ being a constant.
On the other hand, the total energy in this phase is given by
\eqa{E^{*}=\sum_{j=1}^{N} \frac{\left\langle p_{j}^{2}\right\rangle_{*}}{2}+\frac{\kappa}{\tau} \sum_{j=1}^{N}\left\langle 1-\cos \left(\phi_{j}-\phi_{j+1}\right)\right\rangle_{*}.} 
The potential energy and total energy take the following from
\begin{eqnarray}
\la 1-\cos(r_j)\ra_* &=&  (Z^*)^{-1}\int dr (1- \cos (r_j)) e^{-\frac{\kappa}{\tau T^*} (1- \cos (r_j))}  
 \nonumber \\
&=&  1- \frac{I_1(\epsilon)}{I_0(\epsilon)} \nonumber \\
	E^{*}&= & N \frac{T^{*}+\tilde{p}^{2}}{2}+\frac{N \kappa}{\tau} \left( 1 - \frac{I_{1}(\epsilon)}{I_{0}(\epsilon)} \right)
\end{eqnarray}
with  $\epsilon = \kappa /(T^*\tau)$ and $I_{n}(x)$ denotes the modified Bessel's function of order $n$ for the argument $x$. 

The micro-canonical initial energy of the system is $ E_0 = N \frac{\tilde{p}^{2}}{2} + N \frac{\kappa}{\tau}$. Now neglecting the initial growth of the kinetic energy in the transient phase and quasi-conserved nature of kinetic energy in 
prethermal phase, one can consider $E_0  \simeq E^{*} $. One thus arrives at the following  
\eqa{\frac{1}{\epsilon} +  \left( \frac{I_{1}(\epsilon)}{I_{0}(\epsilon)} \right) = 0 \implies \epsilon = 1.06569.} 
This implies $T^* = 0.938363 \frac{\kappa}{\tau}$. In the prethermal phase (with $\cx{p}=0$ ), the average energy is then  given by
\begin{eqnarray}
 E^{*}/N&=& \frac{T^{*}}{2}+\frac{ \kappa}{\tau} \left( 1 - \frac{I_{1}(\epsilon)}{I_{0}(\epsilon)} \right) = \frac{\kappa}{\tau} \Big(\frac{0.938363 }{2} + 0.530819 \Big)\nonumber \\
&=&  \frac{\kappa}{\tau}.
\end{eqnarray}
One can redefine the standard energy per-site, which is equivalent to average energy pumped into the site over a time period. In the prethermal phase, the probability 
distribution of kinetic  and potential energies take the form 
\eqa{ P^*(p) =  \prod_{j^{\prime}=1}^{N} e^{-\left(p_{j^{\prime}}\right)^{2} /\left( 1.876 \frac{\kappa}{\tau}\right)}\\
	P^*(r_j) =  \prod_{j^{\prime}=1}^{N} e^{-\left(1- \cos(r_j)\right) / 1.876 }
}

The specific heat at constant volume can be obtained at the prethermal value as
$C_v = \p E/\p T|_{T^*} $ with 
\begin{eqnarray}
 E &=& \frac{T^{*}}{2}+\frac{ \kappa}{\tau} \left( 1 - \frac{I_{1}( \kappa /(T^*\tau)}{I_{0}( \kappa /(T^*\tau)} \right)   \\
	C_v &=& \p E/\p T |_{T^*} =  \frac{\kappa ^2}{\tau ^2 {T^*}^2}-\frac{\kappa ^2 I_1\left(\frac{\kappa }{{T^*} \tau }\right){}^2}{\tau ^2 {T^*}^2 I_0\left(\frac{\kappa }{{T^*} \tau }\right){}^2} \nonumber \\
	&-& \frac{\kappa  I_1\left(\frac{\kappa }{{T^*} \tau }\right)}{\tau  {T^*} I_0\left(\frac{\kappa }{{T^*} \tau }\right)}+\frac{1}{2} \nonumber \\
	&=&0.88572.
\end{eqnarray}

Note that this value is independent of the driving frequency in the prethermal regime.
It  implies that the system absorbs heat at a constant rate irrespective of the driving frequency in the prethermal state.

The correlation functions assuming the system is in the prethermal GGE can be computed from the sum rule such that area under the correlation
curves keep their area constant, 
$A_K =\sum_x C(x,t) \approx \frac{1}{4} (\la p^4\ra - \la p^2 \ra^2)$, where the average is defined as, $\la y(p) \ra =  \int_{-\infty}^\infty dp ~y(p)~ e^{-p^2/T^*}/ \int_{-\infty}^\infty dp~  e^{-p^2/T^*}$. We find  $ A_K \sim 2{T^*}^2/4 =1.72348 K^2/4 = 0.4402 K^2/\tau^4.$ 
We analyze this  sum rule $A_K$ in the Fig.~\ref{fig:PTcorr} (b) of the main text.  

\section{Exponentially suppressed variation of prethermal temperature and diffusion constant}
\label{diffusion_appendix}

\begin{figure}[h]
	\centering
		\includegraphics[width=1.0\linewidth]{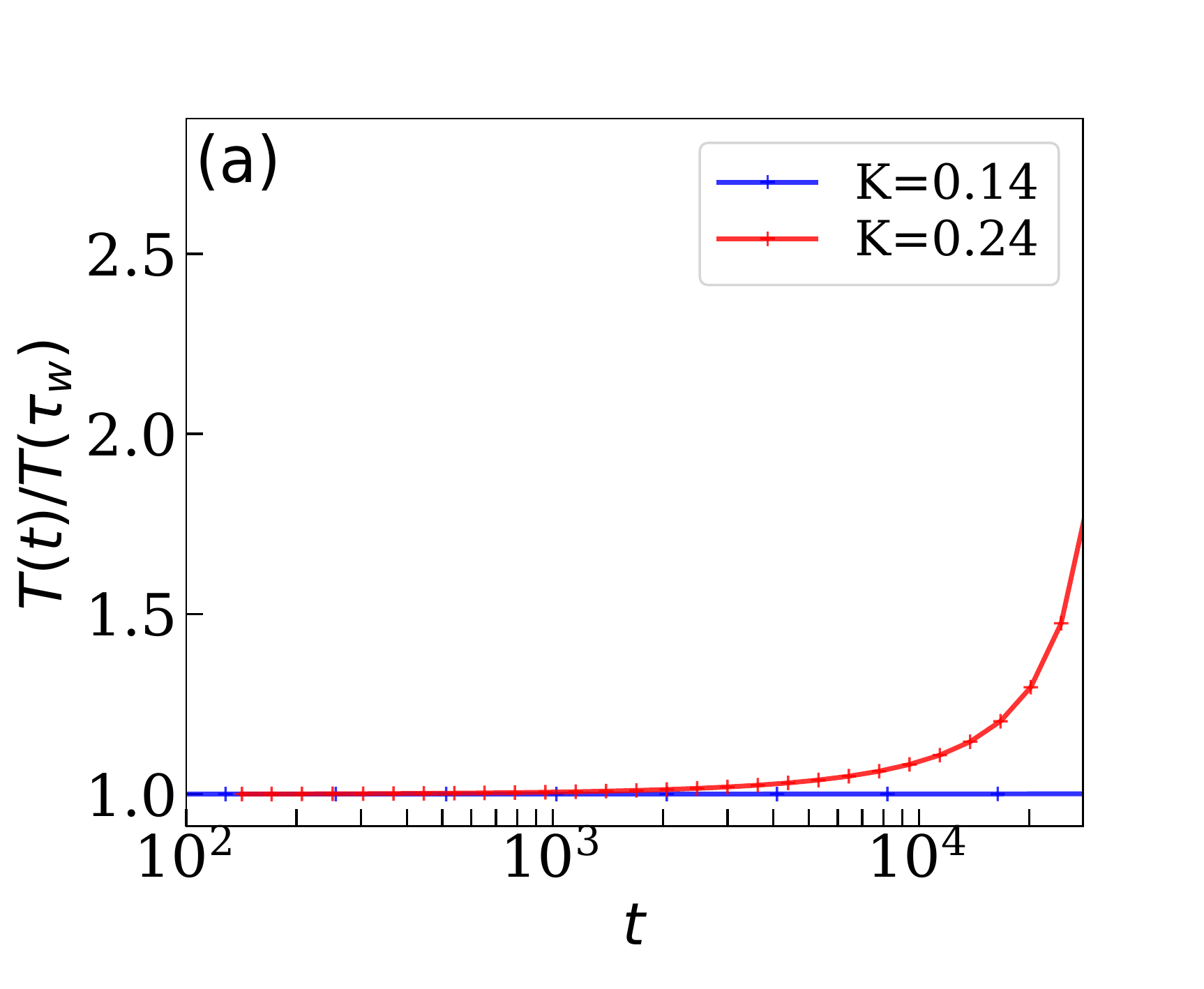}
	\includegraphics[width=1.0\linewidth]{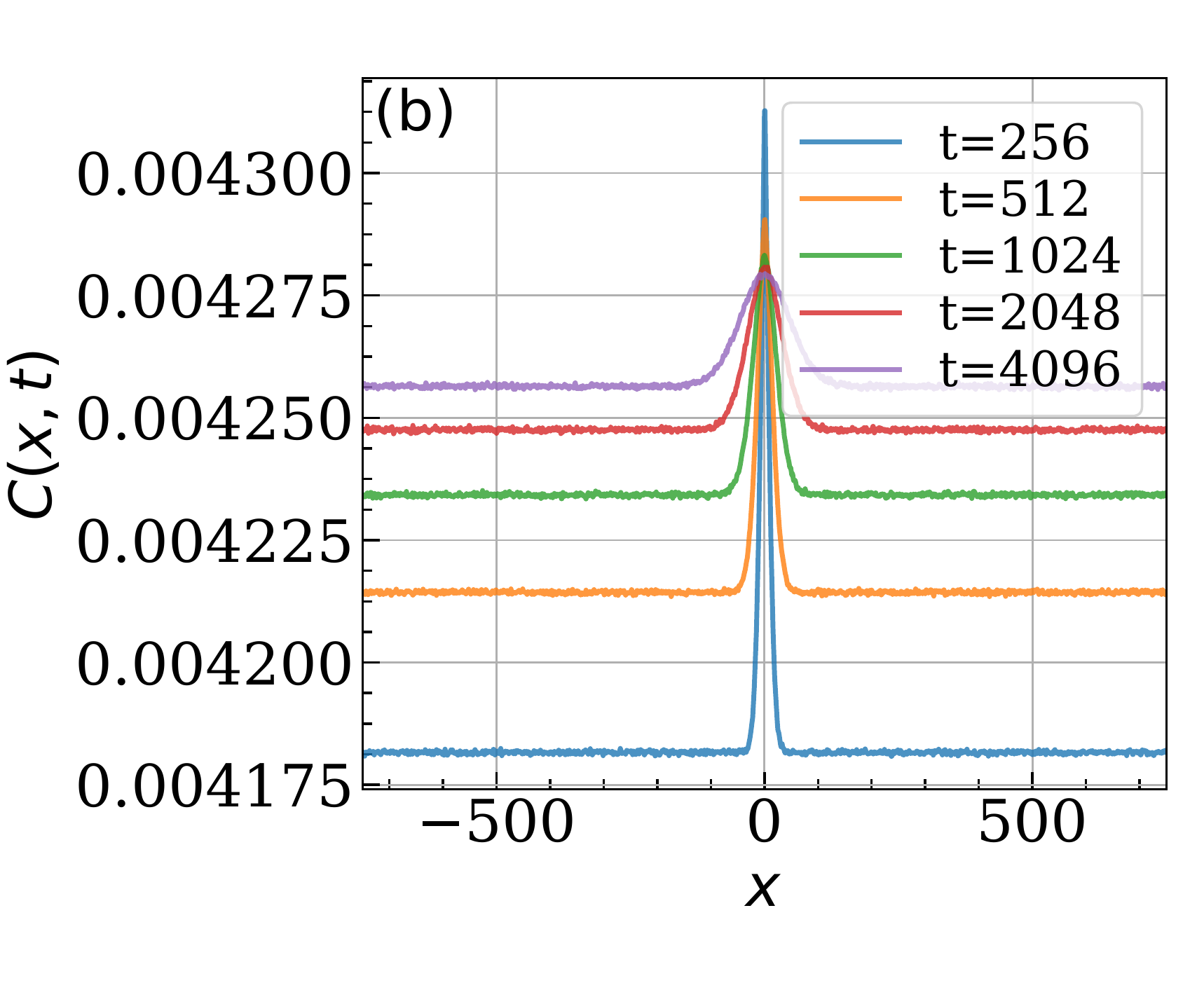}
	\caption{(a)  Plot of prethermal temperature at time $t_w$ to time $t$ as defined by $
	T(t)/T(t_w)=\sum_i p_i(t)^2/\sum_i p_i(\tau_w)^2$. For smaller values $K$, the  heating is suppressed for a larger timescale, while for $K=0.24$ the heating sets in early causing the prethermal temporal window to shrink. (b) Unsubtracted correlations of kinetic energy $\la{p_i^2(t)}{p_j^2(t_w)} \ra$ for $K=0.14$ are shown. We find  the drift in time for the correlation functions. With the deterministic part $ \la {p_i^2(t)} \ra \la {p_j^2(t_w)} \ra$ subtracted, this drift goes to zero. This is in contrast to an equilibrium system, where the average is independent of time. }
	\label{fig:heatingsupp}
\end{figure}

In the prethermal regime, the average kinetic energy is  slowly changing in time, and heating  is exponentially suppressed. This is shown in Fig. \ref{fig:heatingsupp} (a), where the ratio of the  prethermal temperature at time $t$  and $t_w$, defined by $\sum_i p_i(t)^2/\sum_i p_i(\tau_w)^2$,  is analyzed. The almost constant nature of  variable 
$T(t)/T(t_w)$ signifies that the temperature is not significantly changing with time. For high frequency drive with  $K=0.14$, the temporal extent of the prethermal region is larger than that of the for $K=0.24$. These clearly indicate that the departure from the prethermal phase is caused by low frequency drive with higher values of $K$. We present this result in Fig. 1 of the main text.
On the other hand, 
the unsubtracted correlation functions 
$\la{p_i^2(t)}{p_j^2(t_w)} \ra$ in the prethermal region with $K=0.14$ are shown in 
Fig. \ref{fig:heatingsupp} (b) referring to the  drifts for the correlation in time. We note that
correlation propagates even into the  substantially separated rotors from fluctuation core at $x=0$. 
However, once the connected part is subtracted, the correlations decay to zero away from the $x=0$, which suggest the hydrodynamic picture as discussed in the main text.

\begin{figure}[h]
	\centering
	\includegraphics[width=\linewidth]{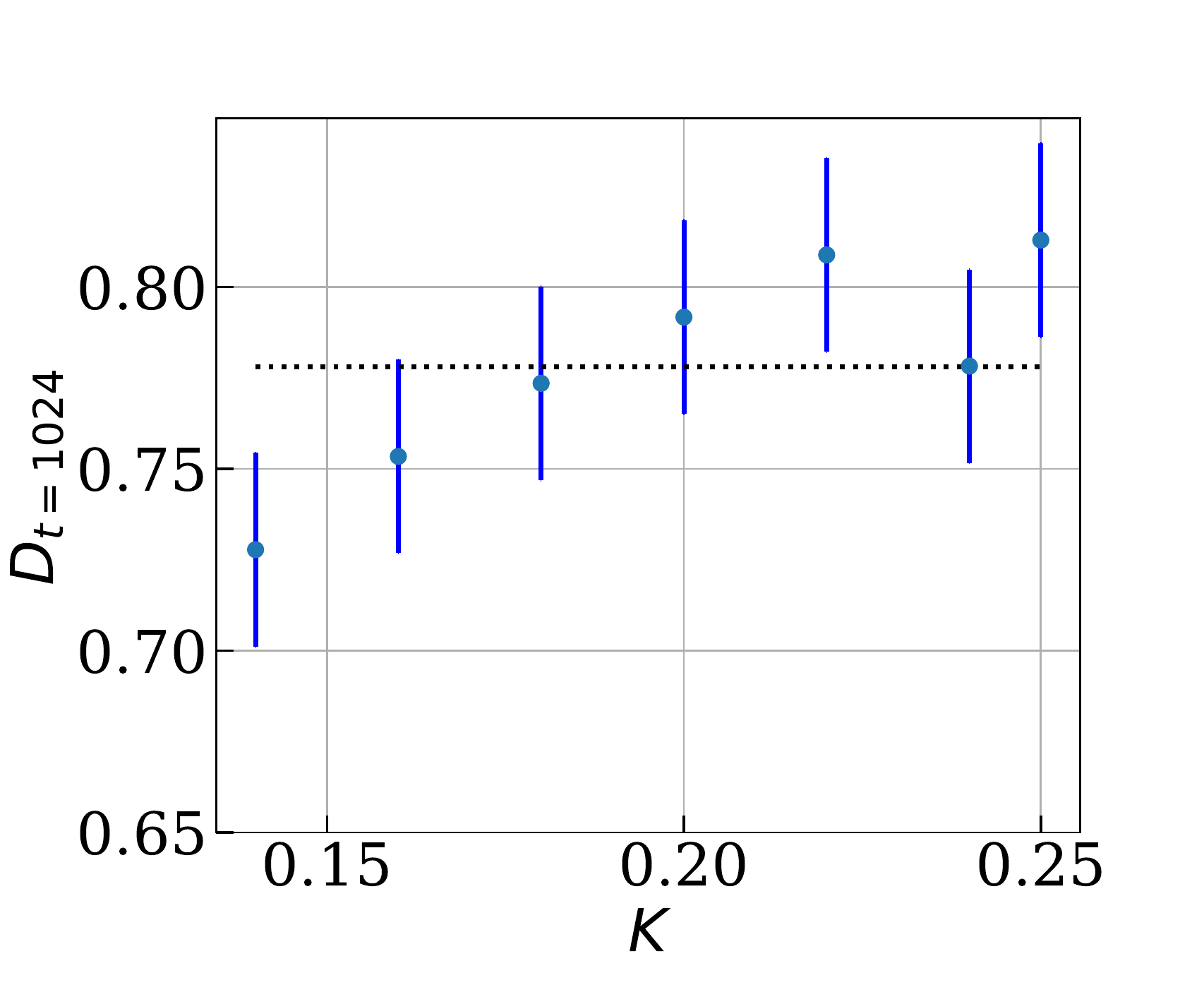}
	\caption{The diffusion constant  extracted by fitting the scaled dynamical correlations at time $t=1024$. The diffusion constant looks very slowly changing with $0.15<K<0.23$ but appears to be saturating around  $0.775$. Note that for smaller $K$, time $1024$ is not large enough for the hydrodynamical scale, hence the results are severely affected by the finite time effect. }
	\label{fig:PTAmpvsK}
\end{figure}

In the prethermal regime, a typical fluctuation spreads in space and time diffusively in the system. A proper diffusive scaling makes the spatio-temporal correlation function collapse to a Gaussian distribution, with two free parameters: the amplitude of the Gaussian and the diffusion coefficient of the Gaussian. The mathematical form of the correlation for a given value of $K$ is thus given by 
 $C(x,t) \sim A_K t^{-1/2} f\big(xt^{-1/2} \big)$, 
with $f(y) = e^{-y^2/2D}/\sqrt{2\pi D}$ where $A_K$ and $D$ are the amplitude and the diffusion coefficient for the Gaussian distribution.  However, extracting the dependence of the diffusion coefficient on driving frequency is numerically more difficult due to both finite time and finite size effects. Our analysis suggests that the diffusion coefficient is very weakly dependent on the driving frequency of the system.  We plot the dependence in Fig. \ref{fig:PTAmpvsK} where it looks like the typical $K$ dependence might be due to finite time effect.

\section{Solution of the hydrodynamic equation}
\label{hydrodynamic_appendix}

We now discuss the hydrodynamic picture of the driven system at times when the prethermal description is valid, i.e. at times much less than the  marginalized time ($t<<t_m$)\cite{rajak2018stability,rajak2019characterizations}. This time scales with the driving frequency as $t_m \sim e^{3/K}$.
The emergent fluctuations in the prethermal region can be  represented by the diffusion equation
\eqa{ \p_t u(x,t) \approxeq \frac{D}{2} \p_x^2 u(x,t)
	+ B \p_x  \zeta (x,t)
	\label{eq1_diffusion}}
 where $u(x,t) = \frac{1}{2} \left(p^2(x,t) - \la p^2(x,t) \ra\right)$ such that $\la u(x,t) \ra \neq \la u(x, t') \ra$.  The  averages are taken over a  GGE at temperature $T^*$. In the prethermal region, the temperature increment  is exponentially suppressed, resulting in an effective hydrodynamic description of the system. The noise $\zeta(x,t)$ of strength $B$ is delta correlated in space and time $\la \zeta(x,t) \zeta(x',t')\ra =  \delta_{xx'} \delta(t-t')$. 
 The derivative under the noise term is due to the almost conserved nature of the total energy of the system in the prethermal phase.  There could be  an additional  exponentially small forcing term present in the diffusion equation. This term might drive the system out of prethermal phase  at later time.
 
 One can solve the above equation by Fourier transform $u(x,t) =  \frac{1}{\sqrt{2\pi}}\int dk e^{i k x} u_k(t)$ and $\zeta(x,t) = \frac{1}{\sqrt{2\pi}} \int dk e^{i k x} \zeta_k(t)$.
 The equation becomes
 \eqa{ \p_t u_k(t) = -\frac{Dk^2}{2} u_k(t) + \iu k B \zeta_k(t),}
 where the noise correlations are now given as $\la \zeta_k(t) \zeta_{k'}(t')\ra =  \delta_{kk'} \delta(t-t')$. 
 The general solution to the above equation is given as,
 \eqa{
u_k(t) = e^{-\frac{Dk^2}{2} t} u_k(0) + \iu k  \int_{-\infty}^t dt' e^{-\frac{Dk^2}{2} (t-t')} B \zeta_k(t') \label{eq:diffsolgen}
}
From this solution the correlation function is obtained by multiplying this equation by $u_k(\tau_w)$ and taking an equilibrium average to get,
\eqa{\la u_k(t) u_{k'}(\tau_w)\ra =  e^{-\frac{Dk^2}{2} (t-\tau_w)} \la u_k(\tau_w)^2 \ra  \delta_{k,k'} }
The average is over the prethermal state.  As  shown in Ref.~\onlinecite{rajak2019characterizations} that the different modes (away from $k=0$) in the prethermal state are independent  and acquires marginal localization independent on the wave numbers. 
Following an inverse Fourier transform, we get the spatio-temporal correlation functions
 as,
\begin{eqnarray}
 C(x,t-\tau_w) & =& \frac{1}{\sqrt{2\pi}} \int dk e^{-\iu k x} e^{-\frac{Dk^2}{2} (t-\tau_w)} \la u_k(\tau_w)^2 \ra \nonumber \\
&=& \frac{ \la u(\tau_w)^2 \ra}{\sqrt{2 \pi D (t - \tau_w)}} e^{-x^2/(2 D (t - \tau_w))}
\end{eqnarray}

 Fluctuation Dissipation
in the large time limit Eq. \ref{eq:diffsolgen}, the first part vanishes and the memory of the  initial condition vanishes. One can obtain 
\eqa{u_k(t) = \lim_{t\to\infty}\iu k  \int_{-\infty}^t dt' e^{-\frac{Dk^2}{2} (t-t')} B \zeta_k(t')}
 Squaring both sides and  using the correlations of the noise, we have
 \begin{eqnarray}
  \la u_k(t)^2 \ra &=&  -k^2  \int_{-\infty}^t dt''  \int_{-\infty}^t dt' e^{-\frac{Dk^2}{2} (t-t')} e^{-\frac{Dk^2}{2} (t-t'')} \nonumber \\
  &\times&B^2 \la \zeta_k(t')  \zeta_k(t'') \ra \nonumber \\
 	&&= -k^2  \int_{-\infty}^t dt''  \int_{-\infty}^t dt' e^{-\frac{Dk^2}{2} (t-t')} e^{-\frac{Dk^2}{2} (t-t'')} 
 	\nonumber \\
 	&\times&B^2  \delta(t'-t'') \nonumber \\
 	&&\overset{t \to \infty} {=}-k^2   \int_{-\infty}^t dt' e^{-\frac{Dk^2}{2} 2(t-t')} B^2 \nonumber  \\
 	&=& ~B^2/D  
 \end{eqnarray}
where we assumed that at large time $t$, the system is approximately in the prethermal state defined by the GGE temperature $T^*$.
The fluctuations of average in the prethermal region is   $\la u_k^2 \ra_{T^*}= \la u^2 \ra_{T^*} = C$.
Note that this relation is exact in the case when the system is static. In the prethermal region,
 this relation is approximately correct to a large timescale, before the heating mechanism takes place.  
 This yields $DC \approxeq B^2$, which we refer to as the approximate fluctuation dissipation relation (AFDR)  in prethermal phase. The correction to this is of the order of $1/k^2$, if we consider a delta correlated noisy forcing term. In the large wavelength limit, i.e. at short but enough coarse-grained space, this contribution goes away and the result approximately holds true.

In the driven system, all the  oscillators individually get kicked periodically in time. The kick inflicts extra energy to the system that gets dissipated in the system eventually. 
This results in an effective  time averaged current in the system $j \sim K \la \nabla_x  V'(x)\ra$. This makes the noise coefficient $B \sim K$. Along with the AFDR, this give that $D \sim K^0$.

\begin{figure}
	\centering
	\includegraphics[width=1.0\linewidth]{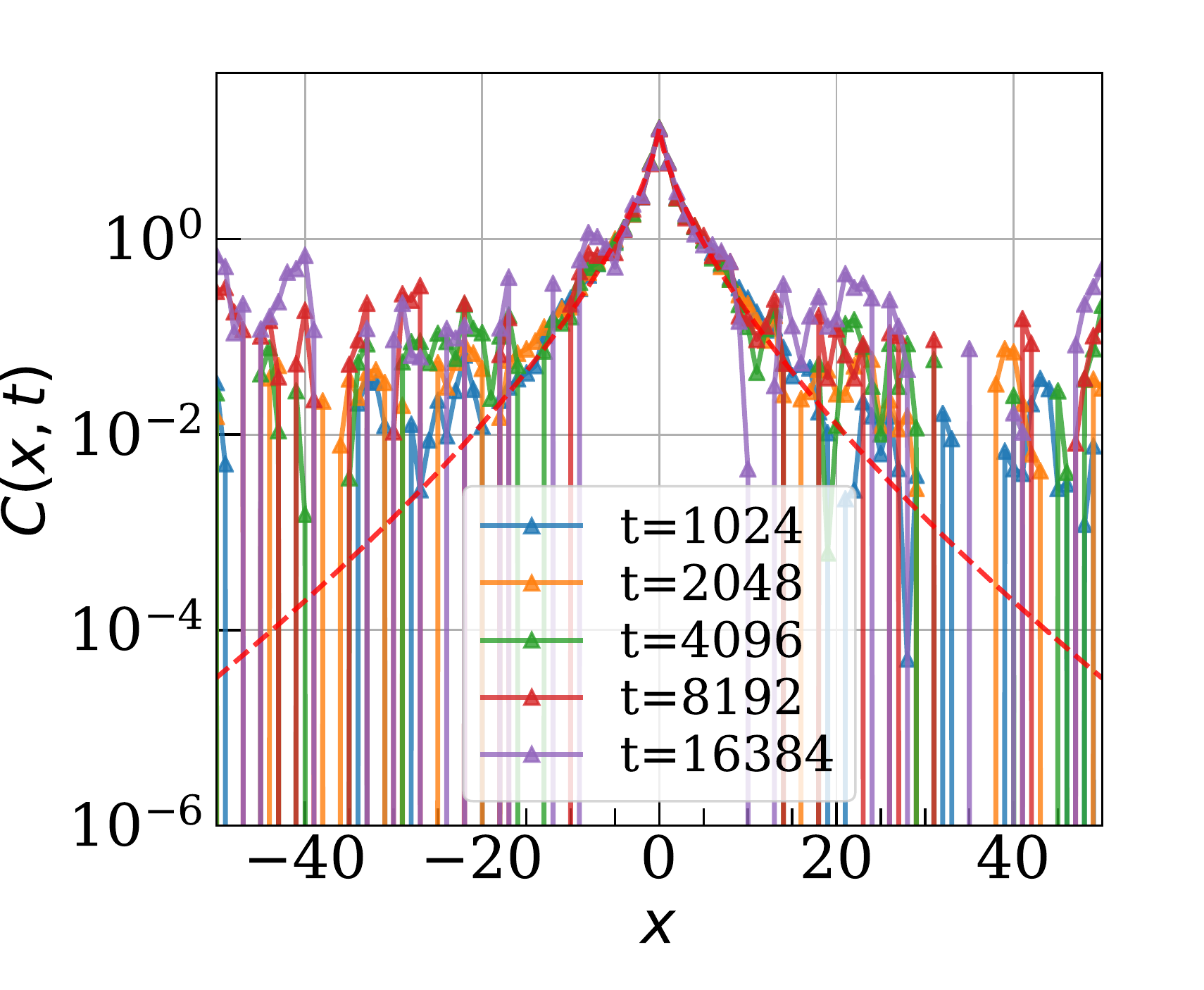}
	\caption{Plot of the correlations in the semi-log scale for the crossover region with $K=0.7$. \textcolor{black}{The same quantity is plotted in Fig.~\ref{fig:IL} in linear scale with stretched exponential fit around the disturbance core at $x=0$. We here emphasize on the accuracy of stretched exponential fit by redrawing it in the semi-log scale. }}
	\label{fig:k0p7log}
\end{figure}

\section{Stretched exponential correlations in crossover region}
\label{crossover_appendix}

We here discuss the intermediate crossover region, where the correlations are exponentially localized in real space with no time dynamics. As mentioned in the main text and Fig.~\ref{fig:IL}  of the main text, we find 
that  the stretched exponential description of the system is in good agreement when the correlation are measured in the close vicinity of fluctuation core $x \ll |N/2|$ where $N$ is the system size $[-N/2:N/2]$.
\textcolor{black}{We redraw the same quantity in semi-log scale with stretched exponential fit in Fig.~\ref{fig:k0p7log}. We can observe that even in logarithmic scale,  the stretched exponential fit of the correlator is appropriate  around the disturbance core $x=0$. This indicates the robustness of the stretched exponential fit of the spatio-temporal correlator in the crossover regime.}
One can clearly notice that correlation becomes heavily  fluctuating with time and space. 
The timescale of this crossover decrease with time, and in infinitely large time limit, the correlations become spatially delta correlated as shown in Fig.~\ref{fig:k1p2} of the main text.

\textcolor{black}{
\section{Frozen correlation in heating region}
\label{heating_analysis}}

\textcolor{black}{
We here discuss the analytical form of the correlation in the heating region that is derived with the help of the assumptions as given in Eq.~(\ref{approx}). We start by 
using Eq.~(\ref{eq_motion}), the momentum of $i$-th rotor at any time 
$t$ can be written as
\begin{equation}
p_i(t)=p_i(0)-K\sum_{n=0}^{t-1}[\sin(r_{i}(n))-\sin(r_{i-1}(n))],
\label{mom_time}
\end{equation}
where $p_i(0)$ is the momentum of the $i$-th rotor at time $t=0$. In our numerical calculations, 
we have assumed the initial momenta of all the rotors to be zero. 
The momentum of a rotor at any stroboscopic time 
$t$ depends on the initial momentum, and the values of angles of that rotor and its nearest neighbors at the earlier time.  
We below derive the correlator assuming the fact that rotors are always uncorrelated from the very beginning of the dynamics. This allows us to write the summation over time starting from $n=0$ as described below. 
However, we know from the numerical analysis that 
driven system requires some time to enter into the heating region. Hence we note that  the independent rotor approximation works better to explain the numerical outcomes
for higher values of $K$ as the driven system enters into the heating region quite early.\\\\
Now the first term of the  kinetic energy fluctuation correlator with $t>t_w$ can be calculated in the heating regime as follows
\begin{align}
&\langle p_i^2(t)p_j^2(t_w)\rangle = 
K^4 \Bigg \langle \sum_{n,n'=0}^{t-1}\sum_{m,m'=0}^{t_w-1}\Bigg[ [\sin(r_{i}(n))-\sin(r_{i-1}(n))]\nonumber \\
&\times [\sin(r_{i}(n'))-\sin(r_{i-1}(n'))] 
\times [\sin(r_{j}(m))-\sin(r_{j-1}(m))] \nonumber \\
& \times 
[\sin(r_{j}(m'))-\sin(r_{j-1}(m'))]\Bigg ]
\Bigg \rangle \nonumber \\
&= K^4 \Bigg \langle \sum_{n,n'=0}^{t-1}\sum_{m,m'=0}^{t_w-1}\Bigg[   
\Big[\sin(r_{i}(n)) \sin(r_{i}(n')) \nonumber \\
&+ \sin(r_{i-1}(n)) \sin(r_{i-1}(n')) \nonumber \\
&- \sin(r_{i}(n)) \sin(r_{i-1}(n')) - \sin(r_{i-1}(n)) \sin(r_{i}(n')) \Big]\nonumber \\
&\times 
\Big [\sin(r_{j}(m)) \sin(r_{j}(m')) + \sin(r_{j-1}(m)) \sin(r_{j-1}(m')) \nonumber \\
&- \sin(r_{j}(m)) \sin(r_{j-1}(m')) - \sin(r_{j-1}(m)) \sin(r_{j}(m')) \Big]
\Bigg ]
\Bigg \rangle \nonumber\\
&= K^4\sum_{n,n'=0}^{t-1}\sum_{m,m'=0}^{t_w-1}\Big( 2~ \delta_{n,m}
\delta_{n',m'} \delta_{i,j} + \frac{1}{2}~ \delta_{n,m} \delta_{n',m'} \delta_{i,j+1} \nonumber \\
&+ \frac{1}{2}~ \delta_{n,m} \delta_{n',m'} \delta_{i,j-1} + \delta_{n,n'}\delta_{m,m'} \delta_{i,i} \delta_{j,j} \nonumber \\
& +\frac{3}{4}~ \delta_{n,n'}
\delta_{m,m'}\delta_{m,n} \delta_{i,i} \delta_{j,j} \delta_{i,j} \nonumber \\
&+ \frac{3}{8}~ \delta_{n,n'}
\delta_{m,m'}\delta_{m,n} \delta_{i,i} \delta_{j,j} \delta_{i,j+1} \nonumber \\
&+ 
\frac{3}{8}~ \delta_{n,n'}
\delta_{m,m'}\delta_{m,n} \delta_{i,i} \delta_{j,j} \delta_{i,j-1}
\Big) \nonumber \\
&=K^4\Big(2~ t_w^2 \delta_{i,j} + \frac{3}{4} t_w  \delta_{i,j}  +  \frac{1}{2}~ t_w^2 \delta_{i,j+1} + \frac{3}{8}~ t_w \delta_{i,j+1} \nonumber \\
&+ \frac{1}{2}~t_w^2 \delta_{i,j-1} + \frac{3}{8}~ t_w \delta_{i,j-1}+ tt_w \delta_{i,i} \delta_{j,j} \Big).
\label{corr1}
\end{align}
\\\\
Here we use the fact that $\delta_{i,j}\delta_{i',j}=\delta_{i,j}\delta_{i,j'}=0$.
The second part $\langle p_i^2(t)\rangle \langle p_j^2(t_w)\rangle$ now looks like 
\begin{align}
&\langle p_i^2(t)\rangle \langle p_j^2(t_w)\rangle = 
K^4 \Bigg \langle \sum_{n,n'=0}^{t-1}
[\sin(r_{i}(n))-\sin(r_{i-1}(n))]\nonumber \\
&\times [\sin(r_{i}(n'))-\sin(r_{i-1}(n'))] 
\Bigg \rangle \nonumber \\
&\times
\Bigg \langle \sum_{m,m'=0}^{t_w-1}
 [\sin(r_{j}(m))-\sin(r_{j-1}(m))] \nonumber \\
& \times 
[\sin(r_{j}(m'))-\sin(r_{j-1}(m'))]
 \Bigg \rangle \nonumber \\
 &= K^4 tt_w \delta_{i,i} \delta_{j,j}
\end{align}
This leads to write the kinetic energy correlator as
\begin{eqnarray}
&&C(i,j,t,t_w)=\frac{1}{4}[\langle p_i^2(t)p_j^2(t_w)\rangle-\langle p_i^2(t)\rangle\langle p_j^2(t_w)\rangle]\nonumber \\
&&= K^4\Big( \frac{t_w^2}{2} \delta_{i,j} + \frac{t_w^2}{8} \delta_{i,j+1} + \frac{t_w^2}{8} \delta_{i,j-1} \Big ) +O(t_w).
\end{eqnarray}
}
\bibliography{ref}

\begin{thebibliography}{113}
\bibitem [{com()}]{computation}%
  \BibitemOpen
  \href@noop {} {}\bibinfo {note} {We compute other fluctuation correlation
  such as, unequal (equal) space and equal (unequal) time correlation, that
  also shows distinct features in the different dynamic phases (in
  preparation).}\BibitemShut {Stop}%
\end{thebibliography}
\end{document}